\documentclass[aps,prl,reprint,superscriptaddress]{revtex4-2}
\usepackage[colorlinks=true,plainpages=false,linkcolor=blue,urlcolor=blue,citecolor=blue,pdfpagemode=UseNone,pdfstartview=FitBH]{hyperref} 
\usepackage[labelformat=empty, justification=justified, singlelinecheck=false]{caption}
\usepackage{ragged2e}
\usepackage{graphicx}% Include figure files
\usepackage{dcolumn}% Align table columns on decimal point
\usepackage{bm}% bold math
\usepackage{xspace}

\newcommand{\cvs}{CsV$_{3}$Sb$_{5}$\xspace}

\newcommand{\avs}{\textit{A}V$_{3}$Sb$_{5}$\xspace}

\newcommand{\fref}[1]{Fig.\,\ref{#1}}
\newcommand{\quadc}{$2\times 2\times 4$\xspace}
\newcommand{\doubc}{$2\times 2\times 2$\xspace}

\newcommand{\epsa}{$\delta \epsilon_a$\xspace}
\newcommand{\bragg}{$(4\bar{2}0)$\xspace}
\newcommand{\vtg}{V $3d$\xspace}
\usepackage{titlesec}
\titleformat{\section}[block]
  {\normalfont\bfseries\raggedright\large}
  {}{0pt}{}

\titlespacing*{\section}{0pt}{0.7\baselineskip}{-0.1\baselineskip}

\titleformat{\subsection}[block]
  {\normalfont\bfseries\raggedright\normalsize}
  {}{0pt}{\leavevmode}

\titlespacing*{\subsection}{0pt}{0.2\baselineskip}{0.2\baselineskip}
\begin{document}

\title{Strain-control of electronic superlattice domains in \cvs}

\author{E.~Sadrollahi}
\email{elaheh.sadrollahi@tu-dresden.de}
\affiliation{Institut f\"ur Festk\"orper- und Materialphysik, Technische Universit\"at Dresden, 01062 Dresden, Germany}

\author{J.~Almeida Mendes}
\affiliation{Institut f\"ur Festk\"orper- und Materialphysik, Technische Universit\"at Dresden, 01062 Dresden, Germany}
\affiliation{Universidade de Coimbra, 3001-401 Coimbra, Portugal}

\author{N.~Stilkerich}
\affiliation{Institut f\"ur Festk\"orper- und Materialphysik, Technische Universit\"at Dresden, 01062 Dresden, Germany}
\affiliation{Max Planck Institute for Chemical Physics of Solids, 01187 Dresden, Germany}

\author{A.\,K.~Sharma}
\affiliation{Max Planck Institute for Chemical Physics of Solids, 01187 Dresden, Germany}

\author{C.~Shekhar}
\affiliation{Max Planck Institute for Chemical Physics of Solids, 01187 Dresden, Germany}

\author{C.~Felser}
\affiliation{Max Planck Institute for Chemical Physics of Solids, 01187 Dresden, Germany}
\affiliation{W\"urzburg-Dresden Cluster of Excellence ct.qmat, Technische Universit\"at Dresden, 01062 Dresden, Germany}

\author{T.~Ritschel}
\affiliation{Institut f\"ur Festk\"orper- und Materialphysik, Technische Universit\"at Dresden, 01062 Dresden, Germany}

\author{J.~Geck}
\email{jochen.geck@tu-dresden.de}
\affiliation{Institut f\"ur Festk\"orper- und Materialphysik, Technische Universit\"at Dresden, 01062 Dresden, Germany}
\affiliation{W\"urzburg-Dresden Cluster of Excellence ct.qmat, Technische Universit\"at Dresden, 01062 Dresden, Germany}

\date{\today}

\begin{abstract}
The kagome metals \avs (A = K, Rb, Cs) provide a unique platform to investigate the physics of interacting electrons, a central challenge in condensed matter physics. A key obstacle in unraveling their correlated behavior is to determine which structural and electronic degrees of freedom are involved and how they couple. Here we address this important issue with a novel approach, namely by exploring the strain dependence of electronic superlattices in \cvs. Using high-resolution x-ray diffraction, we track the detwinning of the \quadc\ electronic crystal and uncover a gigantic strain response of its domains. We further show that the detwinned \quadc\ phase exhibits an intrinsic 2$\textbf{q}$-modulation and strong mode coupling. Density functional theory reveals that the structural \quadc modulation couples strongly to the V $3d$-orbitals, naturally explaining its pronounced strain response. In contrast, the \doubc\ phase at lower temperatures remains essentially unaffected by small uniaxial strain. This dichotomy points to fundamental differences in the symmetry breaking and stabilization mechanisms of the two electronic orders. 

More specifically, our results provide evidence for the active role of orbital degrees of freedom, which can realize distinct, complex ordering patterns driven by competing interactions.

\end{abstract}

\maketitle

\section*{Introduction}

Research on quantum materials is driven by the quest to understand and control their unconventional electronic phases, which not only lie beyond the standard models of condensed matter physics but also hold significant promise for technological applications, particularly in information technology and quantum computing~\cite{Cava:2021aa, Keimer:2017a, Sarma:2015a, Montblanch:2023a, Goyal:2025a}. Precise control of electronic properties at various length scales is fundamental to such technological applications. Indeed, quantum materials with competing or coexisting electronic instabilities offer exceptional electronic tunability and control, paving the way for novel functionalities.

\begin{figure*}[t!]
\centering
\includegraphics[width=\textwidth]{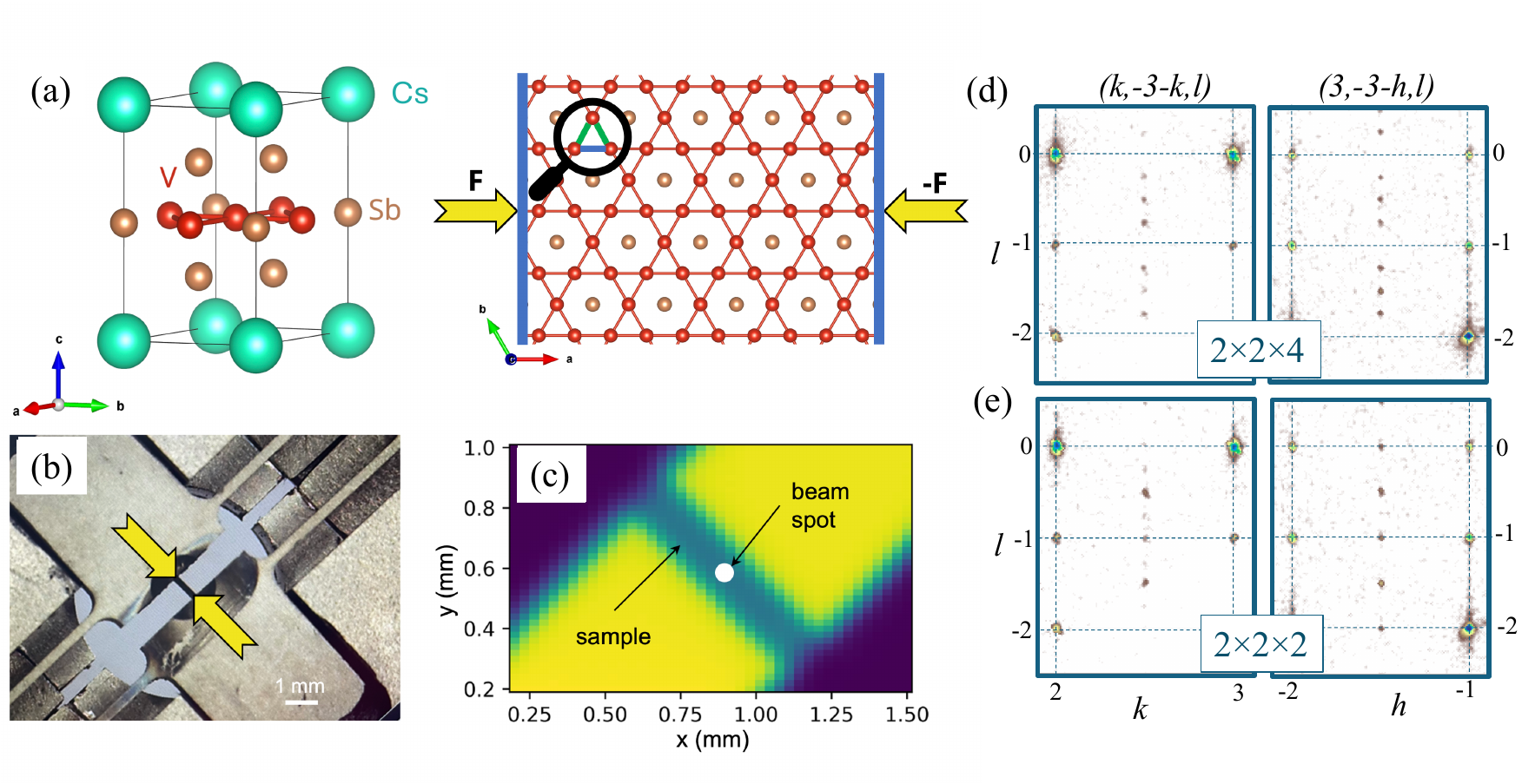}
\caption{\justifying\parfillskip=0pt%
\textbf{Fig. 1 \textbar{} Key elements of the experimental setup for XRD under uniaxial stress.}
(a) Unit cell of the basic P6/mmm lattice structure of \cvs\. (b) Single crystal mounted on a piezo-driven uniaxial strain cell. The arrows indicate the direction of the applied force, which is parallel to the crystallographic $a$-direction. The geometry is also illustrated in (a). Under uniaxial stress, one of the triangular bonds becomes inequivalent to the other two. (c) Absorption scan over the sample region, showing the sample in the center. The size of the beam spot at normal incidence is indicated. The scattering plane, which contains both the incident and diffracted x-ray beams, is oriented horizontally in this panel. (d) XRD intensity maps taken at 80\,K, showing the characteristic superlattice reflections of the \quadc phase. (e) Same intensity maps measured in the \doubc phase at 15\,K.}
\label{intro_figure}
\end{figure*} 

In this context, layered kagome materials, with their two-dimensional networks of corner-sharing triangles, represent particularly compelling examples~\cite{Wang:2024a}. Originally studied for frustrated magnetism in insulators, kagome metals gained prominence when Ortiz \textit{et al.} discovered superconductivity in the material family \avs\ ($A$ = K, Rb, Cs)~\cite{Ortiz:2020h}. Since then, these superconductors have captured the condensed matter community’s attention~\cite{Wilson:2024a}. Beyond their topological and superconducting behavior, these compounds exhibit unconventional electronic orders, which are often labeled a charge density wave (CDW), although their microscopic nature appears to be much more intricate. Indeed, the CDW in \avs\ has been proposed to display (i) self‑organized chirality~\cite{Jiang:2021z}, (ii) time‑reversal symmetry (TRS) breaking~\cite{Mielke:2022aa} (potentially tied to orbital currents~\cite{Kiesel:2013a}), and (iii) an interplay with superconductivity that may form a chiral pair‑density wave state~\cite{Deng:2024aa}.

Despite extensive studies, however, key aspects of the electronic order in \avs remain unresolved. Major points of controversy are whether the electronic order breaks TRS~\cite{Mielke:2022aa, Yu:2021o, Xu:2022a, Jiang:2021z, Farhang:2023aa, Wang:2024a, Huazhou:2022a}, or whether it exhibits chirality~\cite{Jiang:2021z,Elmers:2025}. Similarly, while the $2 \times 2$ translational symmetry breaking in \avs is well established, the precise atomic-scale structure of the electronic ordering remains debated. In \cvs, for example, x-ray diffraction (XRD) studies have proposed staggered star-of-David (SoD) or inverse star-of-David (iSoD) arrangements~\cite{Stahl:2022a}, mixed SoD/iSoD stacking~\cite{Kautzsch:2023a}, or a even more complex phase coexistence~\cite{Xiao:2023aa}. Concerning the crystallographic analysis of XRD data, superlattice (SL) twinning so far severely limits the amount of extractable information, because different structural models can describe the XRD intensities of twinned SLs equally well~\cite{Stahl:2022a,Kautzsch:2023a, Stier:2024b}. This is particularly detrimental for determining the exact origin of the broken six-fold symmetry in the electronically ordered phases as well as for detecting chiral signatures of the electronic order in the bulk. Access to detwinned electronic superlattices therefore represent a major step towards a conclusive resolution of the issues mentioned above.

In this study, we apply small uniaxial compressive strains \epsa along the $a$-direction and investigate the response of the electronic order using high-resolution XRD. In the \quadc phase, even very small compressive strains of the order of -0.1\,\% lead to pronounced changes in the diffraction pattern, highlighting the strong coupling of this electronic order to structural distortions. Our results show that the electronic \quadc superlattice becomes detwinned at \epsa$\simeq -0.12$\,\%, uncovering an intrinsic double-\textbf{q} modulation of this phase. In contrast and quite surprisingly, the \doubc superlattice remains largely unaffected by such strains, demonstrating that the \doubc order couples much weaker to uniaxial strain.

\section*{Results and discussion}
XRD measurements were carried out on \cvs\ single crystals as a function of temperature and strain, as detailed in the Methods section. Key elements of our experimental setup are illustrated in \fref{intro_figure} (a)--(c). Panel (d) and (e) show representative zero-strain XRD data for the mounted sample. All diffraction patterns collected are in excellent agreement with our previously published results~\cite{Stahl:2020aa, Stier:2024b}. 
Note that throughout this work, $h$, $k$, and $l$ refer to the unmodulated $P6/mmm$ structure shown in Fig.\,\ref{intro_figure}\,(a).
We also observe the same temperature-dependent SL: upon cooling, the sample first transitions into the \quadc phase at $T_1 \simeq 95$\,K, and subsequently into the \doubc phase at $T_2 \simeq 60$\,K. Moreover, the resolution-limited Bragg and SL reflections highlight the outstanding quality of the single crystal used.
\begin{figure*}[t!]
\centering
\includegraphics[width=\textwidth]{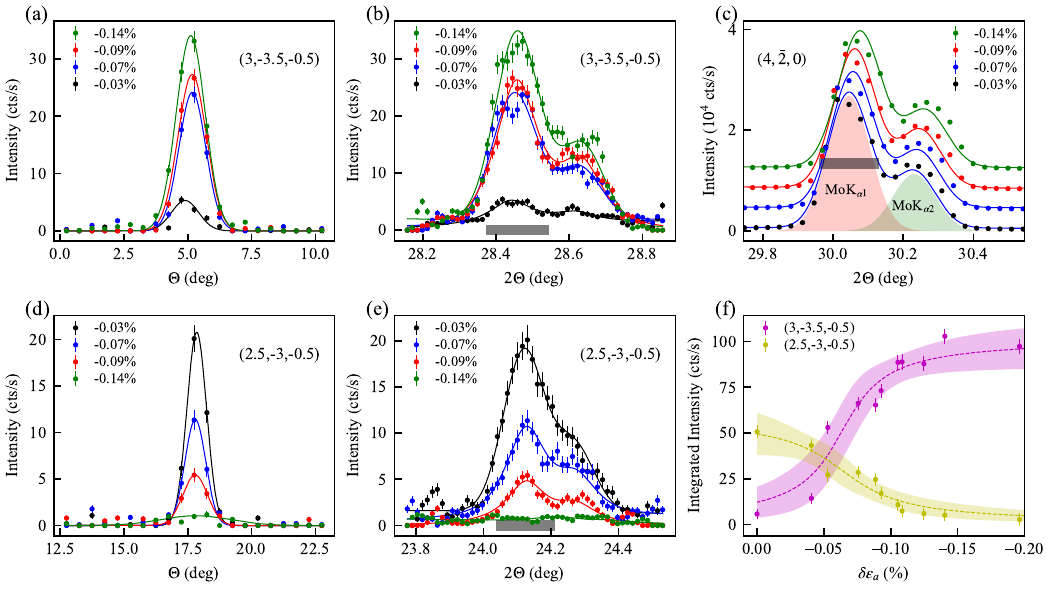}
\caption{\justifying\parfillskip=0pt%
\textbf{Fig. 2 \textbar{} Strain-dependent XRD data taken at 80\,K in the \quadc phase.}
Left panels: $\theta$-scans (rocking scans) through the $(3,-3.5,-0.5)$ and $(2.5,-3,-0.5)$ superlattice reflections as a function of strain.
Middle panels: $\theta 2 \vartheta$-scans (radial scans) of the same superlattice peaks as a function of strain. The width of the adjacent Bragg reflection, which defines the experimental resolution, is indicated by a gray bar. The observed double-peak structure arises from the two wavelengths of the x-ray beam (Mo K$\alpha_{1,2}$).
Right panels: $\theta 2 \vartheta$-scans through the \bragg Bragg reflection as a function of strain. The applied strain is parallel to the \bragg scattering vector. The small shifts in peak position correspond to a strain range of $-0.14$\,\% $\leq$ \epsa $\leq 0$\,\%.
}
\label{fig2}
\end{figure*}

\begin{figure*}[t!]
\centering
\includegraphics[width=\textwidth]{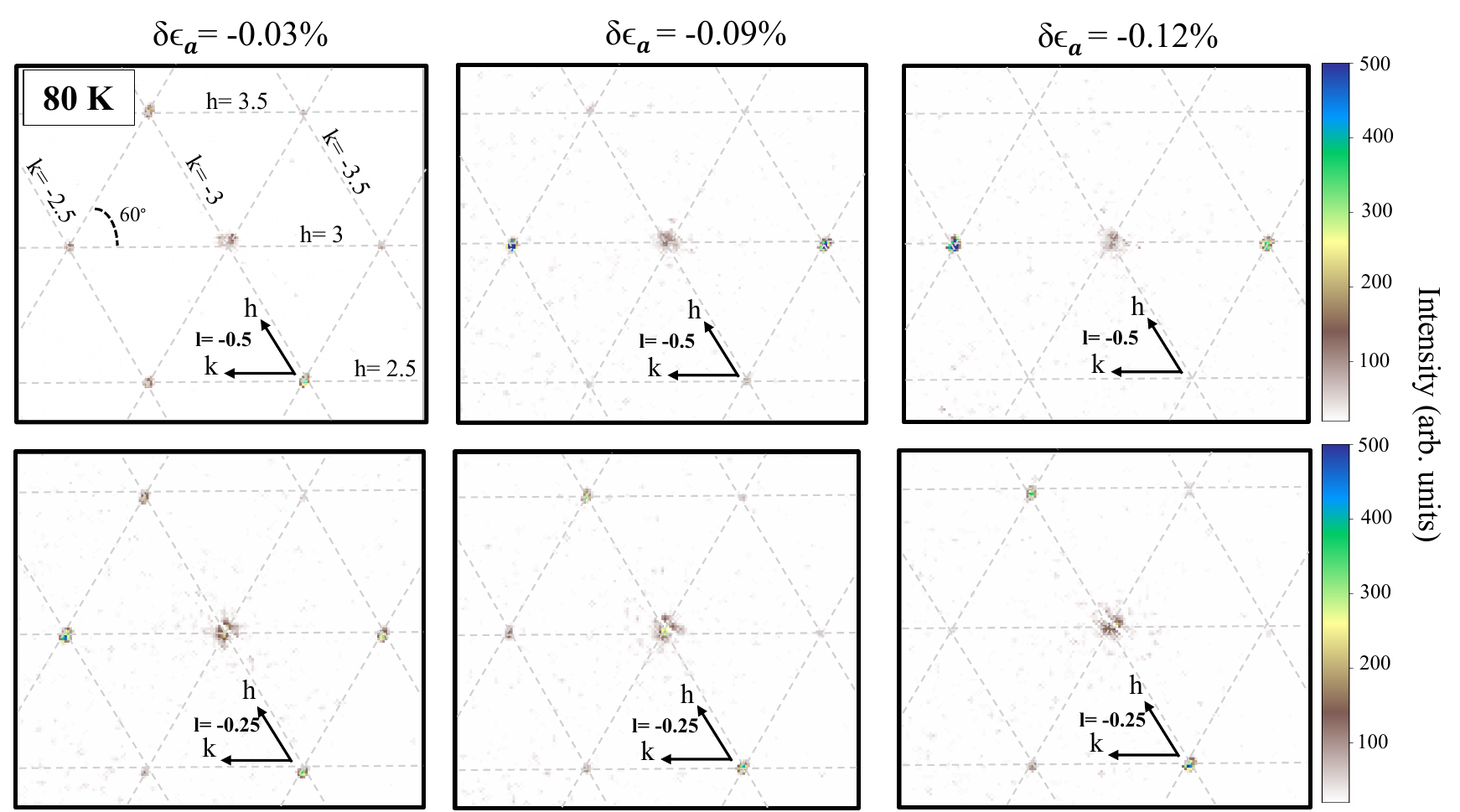}
\caption{\justifying\parfillskip=0pt%
\textbf{Fig. 3 \textbar{} Detwinning of the \quadc superlattice at 80\,K.}
First row: XRD-intensity distribution in the $(h,k,-0.5)$-plane as a function of uniaxial strain. The intensity in the center at $(3,-3,-0.5)$ is due to the neighboring Bragg reflection, while the surrounding reflections are caused by the superlattice modulation. Already a very small strain of only \epsa=$-0.09$\,\% results in a pronounced redistribution of the superlattice peak intensities and at $-0.12$\,\% strain only one pair of superlattice peaks with $h=3$ remains. Second row: Similar measurement, but this time in the $(h,k,-0.25)$-plane. In contrast to the previous case, the superlattice peaks on the $h=3$ line are now suppressed with increasing strain resulting in 4 remaining superlattice peaks at $h=2.5$ and $h=3.5$.}
\label{fig3}
\end{figure*} 

In \fref{fig2}, we present selected line scans as a function of \epsa, measured in the \quadc phase at 80\,K. As seen in this figure, the intensity of the $(3,-3.5,-0.5)$ in (a),(b) and the intensity of the $(2.5,-3,-0.5)$ in (d),(e) exhibit opposite behavior with increasing \epsa: while the intensity of the former increases significantly, the intensity of the latter gradually vanishes. This contrasting behavior is further illustrated in \fref{fig2}\,(f), which displays the integrated intensities of both reflections. Notably, within the resolution of our experiment, the positions and widths of the SL reflections remain unchanged under applied strain. The $\Theta$-scans (rocking scans) in panels (a) and (d) remain sharp, confirming that the mosaicity of the crystal is unaffected by \epsa. Likewise, the $\Theta 2 \Theta$-scans through the SL peaks in panels (b) and (e) remain sharp and resolution-limited, indicating that there is no detectable change in either the correlation length or the modulation vector. From the FWHM of the $\Theta 2 \Theta$-scans we determine the correlation length of the electronic order to be always larger than 27\,nm.

The data shown in \fref{fig2} reveals that different SL reflections can exhibit  very different strain dependences. To clarify the effect of \epsa on the electronic order and the resulting SL modulation, it is therefore necessary to examine a more extended region of reciprocal space. Corresponding data for the \quadc phase are shown in \fref{fig3}, where XRD intensity maps of a section of the reciprocal $hk$-plane at $l = -0.5$ (top) and $l = -0.25$ (bottom) are presented. In these maps, the SL reflections form a regular hexagon around a central intensity feature at $(3, -3, l)$, which originates from the intensity tails of a neighboring Bragg reflection.

As clearly seen in \fref{fig3}, the positions and widths of the SL peaks remain unchanged with increasing \epsa\ at fixed $T$. However, their intensities show a clear and systematic variation: In the top row, the intensity of the six SL peaks at $l = -0.5$ is gradually redistributed, until only one pair of SL reflections with $h=3$ retains significant intensity. This corresponds exactly to the data presented in \fref{fig2}.

Interestingly, the behavior of the SL reflections at $l = -0.25$ is very different. Although a gradual redistribution of intensity is also observed in this case, the two SL peaks at $h=3$ are now suppressed, while the intensity of the remaining four SL reflections increases.
The observed behavior, where SL peaks intensities are redistributed without any change in their position or width, is precisely what one expects for a detwinning of the electronic superlattice: If the electronic order breaks the sixfold symmetry of the underlying $P6/mmm$ lattice, multiple domains with different orientations emerge, resulting in a twinned state.
In XRD, all these differently oriented SL domains contribute to the diffraction pattern, which therefore appears more symmetric than that of a single domain. Applying uniaxial strain along the $a$-axis lifts the symmetry of the $P6/mmm$ lattice, energetically favoring certain domain orientations over others, which enables to alter their population volumes. This leads to a redistribution of SL peak intensities in the XRD pattern --precisely the behavior observed in the \quadc\ phase.

At the same time uniaxial stress may also alter the superlattice modulation within a domain without changing its symmetry and its supercell. However, we argue that this plays no role for the small compressive strains considered: even a small \epsa$=-0.07$\,\% already leads to a pronounced redistribution of the SL-peak intensities (\fref{fig2}). At the same time, this level of lattice strain neither causes any discernible change in the calculated electronic structure\,\cite{Stier:2024b} nor does it create any significant shift in the transition temperature\,\cite{Qian:2021g}. The latter is in fact fully consistent with our observation of $T_1(\epsilon_a)$=constant within the studied strain range (not shown). 
Considering that 0.1\,\% strain correspond to a change of the $a$-lattice parameter of only 0.006\,\AA{} also the underlying lattice structure is not changed significantly.
Since neither the electronic nor the lattice structure are changed by the small \epsa, the stability of the electronic order and the electronic order itself are indeed expected to remain unaffected. 

We therefore conclude that the small strains applied in this study detwin the electronic superlattice without significantly altering the modulation within a twin domain.

Importantly, the data presented in \fref{fig3} indeed reveals that \epsa drives the electronic order into a mono-domain state: As seen in the second row of \fref{fig3}, the diffraction pattern at $l=-0.25$ can be characterized by 2 modulation vectors, namely $\textbf{q}_1=(0.5,0,-0.25)$ and $\textbf{q}_2=(0.5,-0.5, -0.25)$. In principle, these two modulation vectors could still originate from two distinct domains, since uniaxial stress along $\textbf{a}$ only partially lifts the equivalence between the hexagonal directions $\textbf{a}$, $\textbf{b}$, and $\textbf{a}-\textbf{b}$ by making $\textbf{a}$ distinct.  However, the superlattice peaks with $l=-0.5$ in the upper panel can be described by linear combination $\textbf{q}_1+\textbf{q}_2=(1,0.5,-0.5)\equiv (0,0.5,-0.5)$.
This is a crucial observation, as the appearance of superlattice reflections corresponding to linear combinations of the two modulation vectors $\textbf{q}_1$ and $\textbf{q}_2$ necessitates a coherent coupling between the respective modulations --an effect that can only occur if $\textbf{q}_{1,2}$ exist within the \textit{same} volume. 
Our results therefore demonstrate that uniaxial strain along $\textbf{a}$ stabilizes a single superlattice domain and that the superlattice modulation of the \quadc\ is intrinsically a double-\textbf{q} modulation. Furthermore, the observation of superlattice reflections corresponding to $\textbf{q}_1 + \textbf{q}_2$ indicates strong mode coupling~\cite{Fujimoto2015} in full agreement with earlier conclusions~\cite{He2024}. 

\begin{figure*}[t!]
\centering
\includegraphics[width=\textwidth]{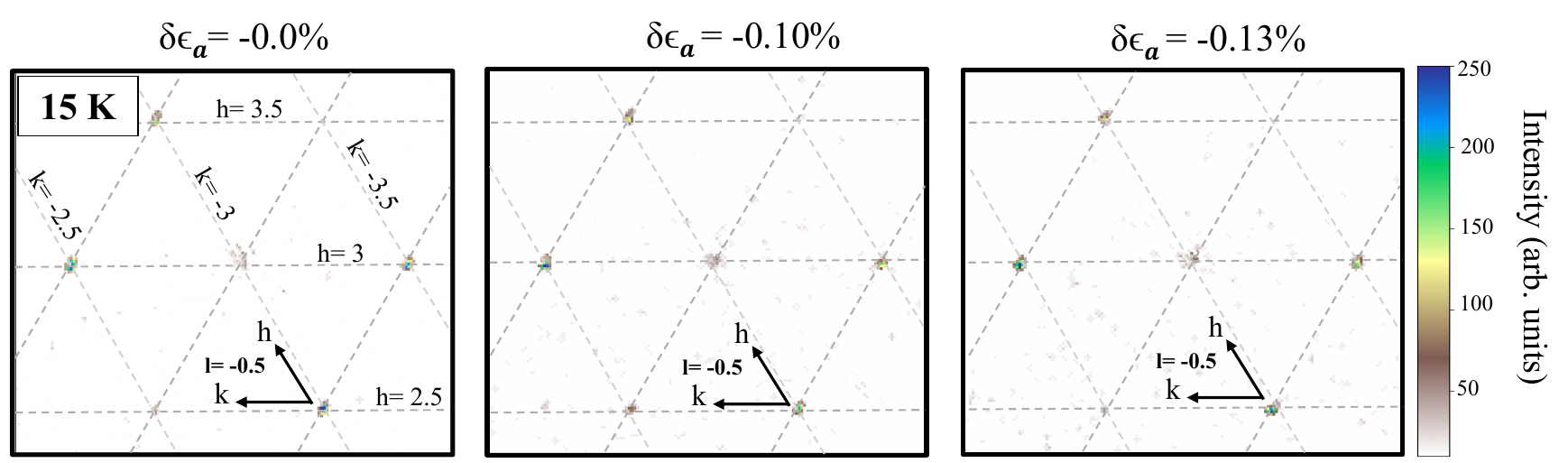}
\caption{\justifying\parfillskip=0pt%
\textbf{Fig. 4 \textbar{} Strain dependence of superlattice reflections in the \doubc phase measured at 15\,K.}
XRD intensity distributions in the $(h,k,-0.5)$ plane are shown. In stark contrast to the data presented in \fref{fig3}, the application of similar strains does not lead to a significant change in the superlattice peak intensities.}
\label{fig4}
\end{figure*}

Another striking observation is the extraordinary sensitivity of the superlattice domains to uniaxial strain. As noted above, a very minute change in the $a$-lattice parameter is sufficient to detwin the \quadc\ phase. This represents a gigantic response, particularly considering that the average crystal structure remains very close to hexagonal; it is solely the weak symmetry breaking by the superlattice which couples so strongly to strain.

\begin{figure}[b!]
\centering
\includegraphics[width=0.85\columnwidth]{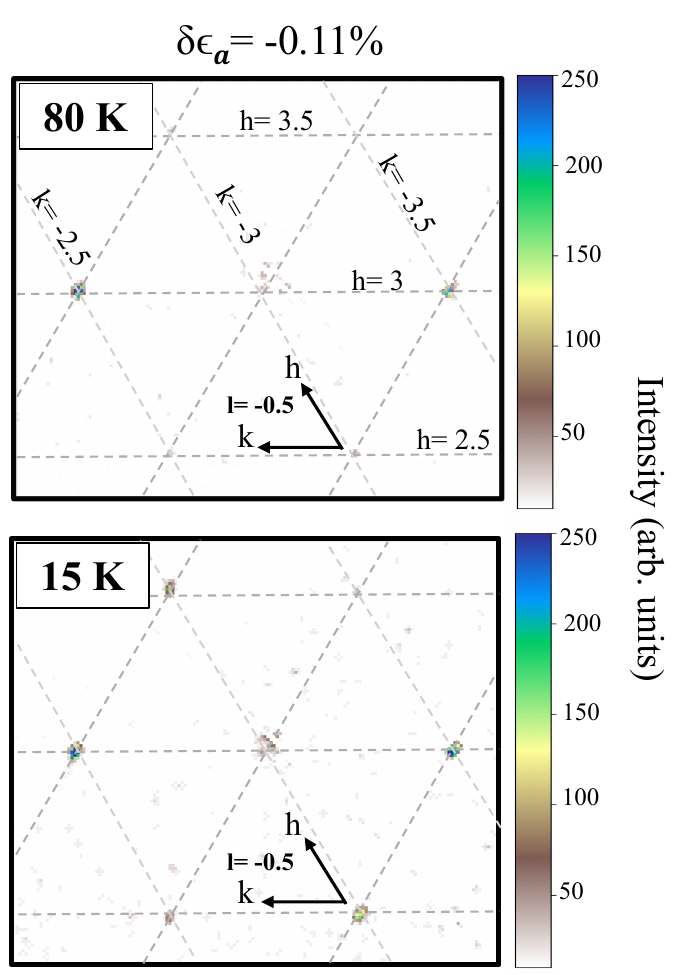}
\caption{\justifying\parfillskip=0pt%
\textbf{Fig. 5 \textbar{} Cooling at constant \epsa=$-0.08$\,\%.}
Upon cooling from room temperature into the \quadc\ phase, the system reaches a detwinned state (80\,K, upper panel). Further cooling into the \doubc\ phase yields an XRD pattern (15\,K, lower panel) that is essentially identical to that shown in \fref{fig4}, despite the fact that the \doubc\ phase was entered from a detwinned state. }
\label{fig5}
\end{figure}

Turning to the strain dependence of the \doubc phase at lower temperatures, surprisingly, we find a completely different response. As shown in \fref{fig4}, cooling the sample into the \doubc phase at \epsa$\simeq0$\,\% and then applying \epsa up to -0.13\,\% does not produce any significant change in the SL peak positions, widths, or intensities. In this sense, the electronic superlattice of the \doubc phase appears to be much more rigid. Even more intriguing are the results shown in \fref{fig5}. In this measurement, a poling strain of \epsa=-0.11\,\% was applied well above $T_1$. The sample was then cooled at constant strain into the \quadc\ phase, resulting in a largely detwinned superlattice, as demonstrated in the top panel of \fref{fig5}. Remarkably, cooling further at \epsa=-0.11\,\% and entering the \doubc phase from this detwinned \quadc state, the resulting SL pattern still closely resembles that of the \textit{unstrained} \doubc state (see \fref{fig3}). These observations imply that the symmetry breaking and, as a consequence, the stabilization mechanism differ fundamentally between the \doubc\ and \quadc\ phases.

Interestingly, inspection of the available structural data~\cite{Stahl:2022a,Kautzsch:2023a, Stier:2024b}  reveals that the bond length disproportionation in the \quadc\ phase is up to about 20 times larger than in the \doubc\ phase ($\sim$2.7\,\% vs. $\sim$0.13\,\%, see supplementary note 1). The deviation from the parent $P6/mmm$ symmetry is therefore much more pronounced in the \quadc\ phase, which aligns very well with the contrasting strain response of the two phases. In order to clarify this dichotomy further, we analyzed the charge density of the \vtg states near the Fermi level using \textit{ab initio} density functional theory (DFT) based on the structural data provided in Refs.\,\onlinecite{Stahl:2022a,Kautzsch:2023a, Stier:2024b}.

\begin{figure*}
    \centering
    \includegraphics[width=2\columnwidth]{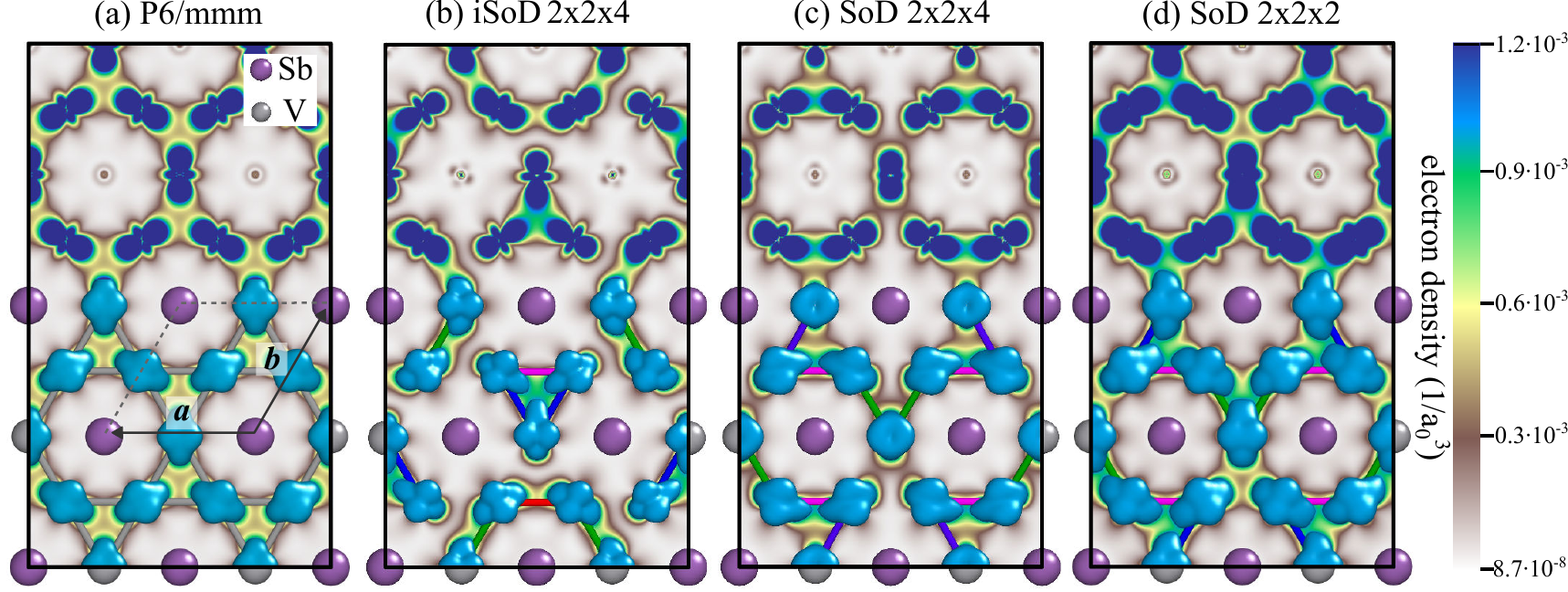}
    \caption{\justifying\parfillskip=0pt%
    \textbf{
    \justifying\parfillskip=0pt%
\textbf{Fig. 6 \textbar{}}
    Density functional theory derived valence electron densities.} 
    (a) Energy-resolved electron density in the V kagome planes, integrated over an energy window from 50\,meV below the Fermi energy up to the Fermi energy, for the undistorted $P6/mmm$ structure. The hexagonal unit-cell vectors are indicated by $\mathbf{a}$ and $\mathbf{b}$. The electron density is visualized within the orthorhombic supercell. V and Sb sites are shown as gray and magenta spheres in the lower half of the supercell, while in the upper half they are omitted to emphasize the false-color plot of the electron density within the kagome plane. An iso-surface corresponding to an electron density value of $10^{-3}$\,$1/a_0^{3}$ ($a_0$: Bohr radius) is displayed in the lower half of the supercell. (b,c) Same as (a), but for the \quadc\ phase structure model from Ref.\onlinecite{Kautzsch:2023a} for the kagome plane at $z=0$ (iSoD distortion) and $z=0.5$ (SoD distortion). Distinct V–V bonds are highlighted in blue, magenta, red and green. This model features three distinct V kagome layers. The third V kagome layer is shown in Supplementary~Note~1. (d) Same as (a,b) and (c), but for the \doubc\ phase structure model from Ref.\onlinecite{Kautzsch:2023a}. The XRD data for the \doubc\ phase can also be refined with a model containing inverse star-of-David (iSoD) clusters, whose corresponding electron density is shown in Supplementary~Note~1.
}
    \label{fig6}
\end{figure*}

As shown in \fref{fig6}, within DFT the \vtg states form intriguing real-space orbital textures. Even in the high-symmetry $P6/mmm$ phase, the \vtg charge density within the kagome layers  shown in panel (a)  closely resembles an ordering of $3z^2-r^2$ orbitals, although the orbitals are in fact linear combinations of different \vtg states with additional structure extending perpendicular to the kagome layers. As demonstrated in panels b) and c), this orbital texture is strongly modified in the presence of structural modulations determined for the \quadc phase. In this phase, a more complex order with dimers and trimers appears, highlighting the importance of electron–lattice interactions in this state. 
By contrast, the real-space orbital texture shown in panel d) for the \doubc phase, which exhibits an about twenty-fold smaller bond-length disproportionation, remains much closer to that of the high-symmetry phase, as expected. This further supports the notion that the two electronic orders are stabilized by different driving mechanisms.

Using ISODISTORT~\cite{Campbell2006, Stokes_ISODISTORT} to determine the isotropy subgroups of the parent space group $P6/mmm$, we find that the two modulation vectors $\textbf{q}_{1,2}$ identified above are consistent with an orthorhombic \quadc\ SL, in agreement with earlier findings~\cite{Stahl:2022a,Kautzsch:2023a, Stier:2024b}. This further supports the intrinsic 2$\textbf{q}$-modulation of the \quadc\ phase and indicates that the \quadc\ order is driven by an instability at the $U$-point of the hexagonal Brillouin zone. The \doubc corresponds to an instability at the $L$-point. Interestingly, our isotropy subgroup analysis shows that in this case all candidate space groups retain mirror symmetries, thereby excluding a chiral electronic \doubc\ order in the bulk. We note, however, that a chiral state in the bulk can still emerge under an applied electric field or at a surface where the symmetry of the bulk is naturally broken.

To conclude, we have presented a temperature- and strain-dependent XRD study of the superlattice modulations in \cvs. Our results demonstrate that the electronic superlattice of the \quadc phase can be detwinned in a well-controlled manner by applying small compressive strains. Remarkably, tiny changes of the lattice parameter $a$ of less than 0.006\,\AA\ are sufficient to induce a pronounced response of the SL domains. This does not contradict our earlier conclusion that such small strains do not significantly affect the SL modulation within each domain: the energy scales and processes governing the formation of the SL itself and the reconfiguration of the SL domain structure are fundamentally different. For example, the domains stabilized by \epsa likely grow at the expense of differently oriented domains through domain wall motion, which is distinct from the process of the SL formation itself. If domain wall motion indeed dominates, the pronounced response of the \quadc domains suggests that the domain walls in \cvs are highly mobile.

Importantly, the detwinning via small \epsa paves the way for a much more precise refinement of the structural modulation in the \quadc phase. Such a refinement is expected to uniquely establish the space group and atomic-scale structure of the \quadc phase, thereby providing decisive insight into the nature of the bulk electronic order and, in particular, determining whether this phase can be chiral. Achieving this goal, however, requires measurements over a much broader region of reciprocal space than was accessible in the present experiment, and will therefore be the focus of future work. But even without detailed structural refinement, our data clearly demonstrate that the electronic order in the \quadc phase is intrinsically a double-\textbf{q} state. The presence of superlattice reflections arising from the coupling of two distinct modulations further points to mode coupling and strong anharmonic lattice effects, consistent with earlier reports~\cite{He2024}.

Remarkably, the pronounced strain sensitivity of the \quadc phase stands in stark contrast to the behavior of the \doubc phase, which remains essentially unaffected within the same range of applied \epsa: The XRD pattern of the \doubc phase shows no significant change in either intensity, position or width of the superlattice reflections. While the precise origin of this contrasting behavior remains to be established, it implies a fundamental difference in the symmetry breaking and stabilization of the two electronic orders. Indeed, the bond-length disproportionation in the \quadc phase is about ten times larger than in the \doubc phase, providing a natural explanation for their contrasting strain responses. Furthermore, our DFT studies based on experimentally determined structures provide strong evidence for an active role of the \vtg orbital degrees of freedom in \cvs, as both the \quadc and \doubc phases exhibit complex but distinct orbital textures in real-space. Nonetheless, the role of orbital degrees of freedom in the \avs kagome superconductors clearly warrants further investigation. In particular, it will be interesting to determine whether and how these complex orbital textures can give rise to the previously reported signatures of chirality and time-reversal symmetry breaking.

\section*{Methods}
\subsection*{Samples}

The \cvs single crystals used in this study are identical to those employed in a previous work~\cite{Stahl:2020aa}. They were synthesized via the self-flux method and thoroughly characterized prior to the experiments. From these crystals, a single-crystalline rod with dimensions $165\,\mu\mathrm{m}\times 55\,\mu\mathrm{m} \times 1300\,\mu\mathrm{m}$ was prepared, with its long axis aligned along the hexagonal $a$-axis. This rod was mounted in Razorbill CS200T piezoelectric-driven uniaxial stress cell and pressurized along the $a$-direction. The entire assembly was placed inside a helium flow cryostat, as shown in Figs.\,\ref{intro_figure}\,(a)and (b). 

\subsection*{X-ray diffraction (XRD)}
XRD measurements as a function of temperature and strain were performed using our custom-built laboratory instrument VEGA, which is optimized for both high resolution and sensitivity. The setup features a monochromatized Mo K$_{\alpha1,2}$ source with a beam diameter of $70\,\mu\mathrm{m}$ at the sample position, and a 300\,K CdTe area detector with no readout noise, offering high detection efficiency and low background. The sample was cooled from 300\,K down to 10\,K using a low-vibration pulse-tube cryostat mounted on a high-precision four-circle diffractometer. The cryostat was equipped with a piezo-driven strain cell specifically designed for transmission XRD experiments, allowing the strain state of the sample to be controlled \textit{in-situ} during cooling. Single crystal datasets were recorded in shutter-less mode where the sample is continuously rotated while the detector is recording frames. Reciprocal space maps were derived from the single crystal datasets using the CrysalisPRO software suite~\cite{crysalisPRO}. For the peakshape and intensity analysis we used custom Python scripts based on the PyFAI~\cite{pyFAI} and FabIO~\cite{fabIO} libraries.

\subsection*{Density functional theory}
The first-principle calculations have been done using the FPLO package (version 22)~\cite{Koepernik1999}.
We used the local density approximation (LDA) of the exchange-correlation potential~\cite{Perdew1992}. The total density was converged on a grid of $12\times12\times12$ and $8\times8\times1$ k-points for the P6/mmm normal cell and the  super cell calculations, respectively. Brillouin zone integration was done using the Mathfessel-Paxton integration scheme~\cite{Methfessel1989}. All calculations were performed using the structural models for the \quadc and \doubc phase from Ref.~\onlinecite{Kautzsch:2023a}. 

\subsection*{Strain determination}
The displacement resulting from the applied force was monitored using a calibrated capacitor, as described in Ref.~\cite{Ikhlas:2020a}. However, the capacitor measures the total displacement of the strain cell, which includes not only the compression of the sample but also contributions from other components such as the epoxy and the sample carrier. It therefore does not directly represent the actual strain in the sample. A distinct advantage of the present XRD setup is its ability to determine the strain variation \epsa directly within the probed sample volume at the center of the sample rod, where the strain is homogeneous. This is illustrated in Fig.\,\ref{intro_figure}\,(c). 

All \epsa-values reported in the following were obtained by fitting the intensity profiles of $\Theta 2 \Theta$-scans (radial scans) through the \bragg-reflection (cf. \fref{fig2}). The distance of the \bragg lattice planes is parallel to the $a$-direction in direct space, hence allowing to monitor the strain variations \epsa directly and with high accuracy of $\pm0.01$\,\%. While the accuracy of the \epsa-measurement is high, we cannot precisely determine the absolute value of $\epsilon_a=\epsilon_a^{(0)}+\delta \epsilon_{a}$, because the zero strain condition cannot be determined directly from our data. However, comparing the superlattice (SL) intensities and the $a$-lattice parameter at different voltages to data measured under zero-strain conditions~\cite{Stahl:2020aa}, we estimate the offset to be $\epsilon_a^{(0)}=-0.02$\,\%, which is very small. In this study, we will therefore refer to \epsa only, which is the quantity determined directly by the present experiment.

\section*{Data availability}
The source data for all of the data and the figures in this work are available from the corresponding author upon request.

\section*{Acknowledgments}
This research has been supported by the Deutsche Forschungsgemeinschaft through SFB 1143 (project-id 247310070), the W\"urzburg-Dresden Cluster of Excellence on Complexity and Topology in Quantum Matter ct.qmat (EXC 2147, project-id 390858490).  We also gratefully acknowledge the support provided by the DRESDEN-concept alliance of research institutions. The authors further thank Dirk Samberg and Fabian Stier for technical support, as well as Noreen Damme for administrative assistance.

\section*{Author contributions}
A.K.S., C.S., and C.F. synthesized and provided the \cvs single crystals. E.S. and N.S. prepared the samples for the strain cell, including crystal orientation, polishing, and cutting. E.S., T.R., N.S., and J.G. conceived the XRD measurements under strain. T.R., J.A.M., E.S., and J.G. analyzed the data and interpreted the results. T.R. performed the DFT study. J.G. and E.S. wrote the manuscript with input from all authors. J.G. initiated and supervised the project.

\end{document}

\title{Supplementary Information \\ Contrasting strain response electronic superlattices in \cvs}
\onecolumngrid

\setcounter{figure}{0}
\renewcommand{\figurename}{FIG.}
\renewcommand{\thefigure}{\arabic{figure}}

\setcounter{table}{0}
\renewcommand{\tablename}{Table}
\renewcommand{\thetable}{\arabic{table}}

\renewcommand\floatpagefraction{.99}

\author{E.~Sadrollahi}
\email{elaheh.sadrollahi@tu-dresden.de}
\affiliation{Institut f\"ur Festk\"orper- und Materialphysik, Technische Universit\"at Dresden, 01062 Dresden, Germany}

\author{J.~Almeida Mendes}
\affiliation{Institut f\"ur Festk\"orper- und Materialphysik, Technische Universit\"at Dresden, 01062 Dresden, Germany}
\affiliation{Universidade de Coimbra, 3001-401 Coimbra, Portugal}

\author{N.~Stilkerich}
\affiliation{Institut f\"ur Festk\"orper- und Materialphysik, Technische Universit\"at Dresden, 01062 Dresden, Germany}
\affiliation{Max Planck Institute for Chemical Physics of Solids, Dresden, Germany}

\author{C. Shekhar}
\affiliation{Max Planck Institute for Chemical Physics of Solids, Dresden, Germany}

\author{C. Felser}
\affiliation{Max Planck Institute for Chemical Physics of Solids, Dresden, Germany}
\affiliation{W\"urzburg-Dresden Cluster of Excellence ct.qmat, Technische Universit\"at Dresden, 01062 Dresden, Germany}

\author{T.~Ritschel}
\affiliation{Institut f\"ur Festk\"orper- und Materialphysik, Technische Universit\"at Dresden, 01062 Dresden, Germany}

\author{J.~Geck}
\email{jochen.geck@tu-dresden.de}
\affiliation{Institut f\"ur Festk\"orper- und Materialphysik, Technische Universit\"at Dresden, 01062 Dresden, Germany}
\affiliation{W\"urzburg-Dresden Cluster of Excellence ct.qmat, Technische Universit\"at Dresden, 01062 Dresden, Germany}

\date{\today}
\begin{abstract}
    
\end{abstract}

{
%\let\clearpage\relax
%\maketitle
}
\begin{center}
    \Large Supplementary Information \\ Strain-control of electronic superlattice domains in \cvs
\end{center}

\section*{Supplementary Note 1: Density functional theory}
\begin{figure}[bh]
    \centering
    \includegraphics[width=\textwidth]{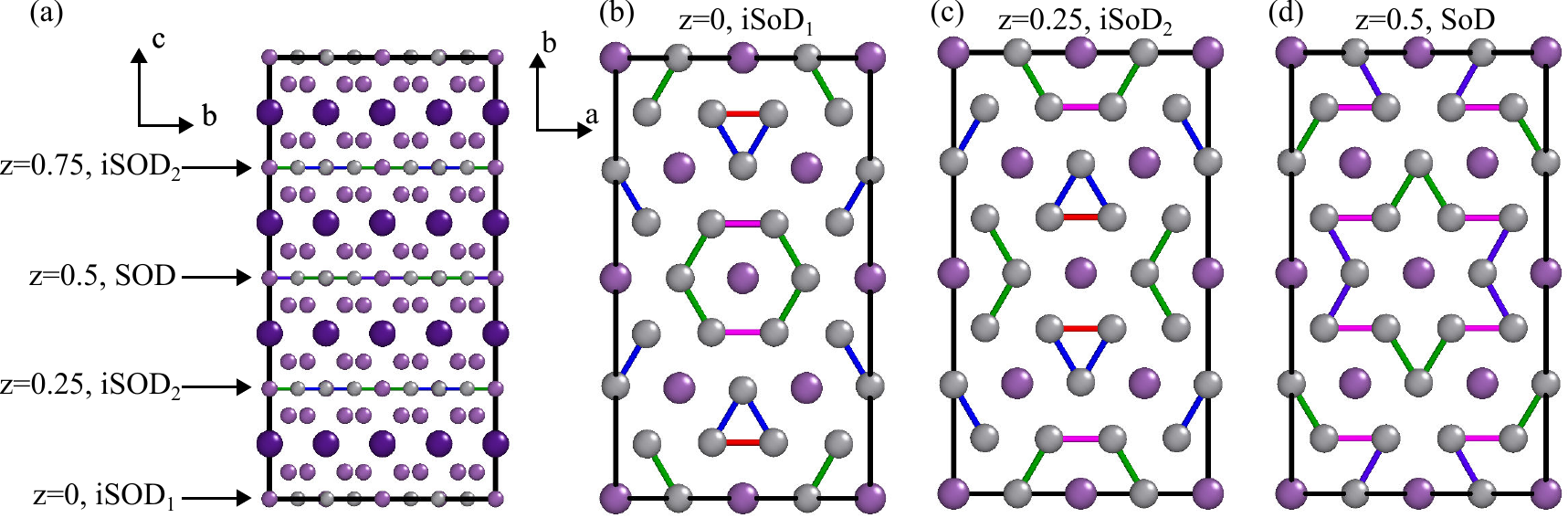}
    \caption{Star-of-David (SOD) and inverse star-of-David (iSOD) distortion within the V kagome planes of the structure model for the \quadc phase from Ref.~\onlinecite{Kautzsch:2023a}. (a) side view on the \quadc structure. One \quadc supercell contains three distinct V kagome planes labeled as SOD, iSOD$_1$ and iSOD$_2$. (b-d) top view of the three V kagome planes. Distinct V-V bonds are shown in green, blue, red and magenta. The V-V bond lengths are listed in \tab~\ref{stab1} along with those of the \doubc phase.}
    \label{sfig1}
\end{figure}

In Ref.~\onlinecite{Kautzsch:2023a} structure models for the \quadc and \doubc phases were derived based on x-ray diffraction data. In \fig~\ref{sfig1}~(a-d) we show a visualization of the \quadc structure model. It contains three distinct V kagome layers: Two at $z=0$ and $z=\pm0.25$ with an inverse star-of-David (iSOD) distortion and one at $z=0.5$ with a star-of-david (SOD) distortion. 

\begin{figure}
    \centering
    \includegraphics[width=\textwidth]{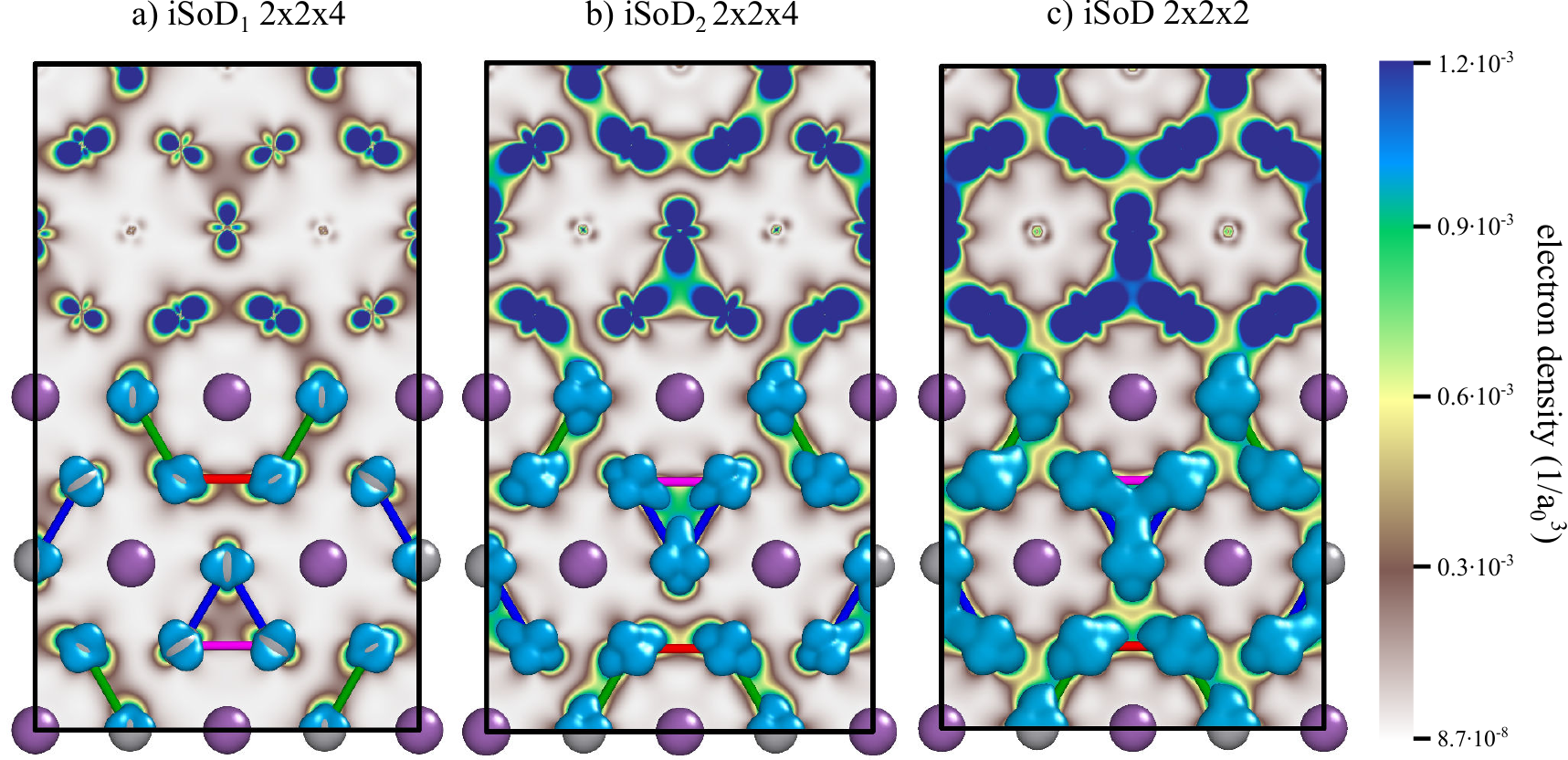}
    \caption{Energy resloved electron density derived from density functional theory. (a,b) Energy resolved electron density corresponding to an energy window from 50\,meV below the Fermi energy to the Fermi energy for the structure model for the \quadc phase from Ref. \onlinecite{Kautzsch:2023a} for the V kagome planes at $z=0$ and $z=0.25$, respectively.. (c) The same as (a,b) but for the structure model for the \quadc phase from Ref. \onlinecite{Kautzsch:2023a}. Distinct V-V bonds are shown in blue, magenta and green. See also Fig.~6 of the main text and \fig~\ref{sfig1}.}
    \label{sfig2}
\end{figure}

As shown in Refs.~\onlinecite{Stahl:2022a,Kautzsch:2023a,Stier:2024b} the XRD data for the \doubc phase can be equally well refined using a structure model containing either SOD distortions or iSOD distortions.
In \tab~\ref{stab1} we summarize the V-V bond lengths derived from the structure models for the distorted kagome planes for both the \quadc and \doubc phase. We also list the bond disproportionation within the characteristic structural motifs, i.e. triangles, hexagons and star-of-David clusters. Our first principle density functional calculations were based on these structure models. Fig.~6 of the main text shows the electron density corresponding to the layers at $z=0.25$ (iSOD$_2$) and $z=0.5$ (SOD). The electron density corresponding to the remaining iSOD$_1$ and the iSOD plane of the \doubc phase is presented in \fig~\ref{sfig2}.

% Command to draw a colored dash with optional rotation
\newcommand{\colordash}[2][]{%
  \tikz[baseline=-0.5ex]\draw[line width=1.5pt, #2, rotate=#1] (0,0) -- (0.3,0);%
}
\newcommand{\mytr}{%
    \tikz{\draw[black] (-30:2mm) -- (90:2mm) -- (210:2mm) -- (-30:2mm);}
}
\newcommand{\myhex}{%
    \tikz{\draw[black] (0:2mm) -- (60:2mm) -- (120:2mm) -- (180:2mm) -- (240:2mm) -- (300:2mm) -- (360:2mm);}
}

\newcommand{\mysod}{%
    \tikz{\draw[black] (0:1.15mm) -- (30:2mm) -- (60:1.15mm) -- (90:2mm) -- (120:1.15mm) -- (150:2mm) -- (180:1.15mm) -- (210:2mm) -- (240:1.15mm) -- (270:2mm) -- (300:1.15mm) -- (330:2mm) -- (360:1.15mm);}
}

\begin{table*}[h]
\centering
\renewcommand{\arraystretch}{1.5}
\begin{ruledtabular}
\caption{Selected V-V bond lengths within the V kagome planes in the \quadc and \doubc phase. The last three rows show the bond disproportionation within the iSOD (triangle and hexagon) and the SOD distortions.}
\begin{tabular}{clllllll}
%\textbf{V--V bond length in \AA} & & & \\
 & & \multicolumn{3}{c}{\quadc} & \multicolumn{2}{c}{\doubc} \\
  & & iSOD$_1$ & iSOD$_2$ & SOD & iSOD & SOD\\
 \hline
\colordash[60]{blue} & $d$ (\AA)  & 2.611(8)  & 2.670(8)  & 2.721(6)  & 2.726(19) & 2.7381(17) \\
\colordash[0]{red}   &   $d$ (\AA) & 2.646(9)  & 2.694(9)  & -- & 2.725(3) & -- \\    
\colordash[-60]{green}&  $d$ (\AA) & 2.710(7)  & 2.724(5)  & 2.746(10) & 2.7501(16) & 2.7388(19) \\
\colordash[0]{magenta}&  $d$ (\AA) & 2.714(16) & 2.736(9)  & 2.673(7) & 2.744(3) & 2.7352(16) \\
\hline 
\mytr & $(d_\mathrm{max} - d_\mathrm{min})/d_\mathrm{min}$ (\%)  & 1.3 & 0.89 & -- & 0.04 &  -- \\
\myhex & $(d_\mathrm{max} - d_\mathrm{min})/d_\mathrm{min}$ (\%)  & 0.11 & 0.44 & -- & 0.22 & -- \\
\mysod & $(d_\mathrm{max} - d_\mathrm{min})/d_\mathrm{min}$ (\%) & -- & -- & 2.7 & -- & 0.14\\
\end{tabular}
\label{stab1}
\end{ruledtabular}
\end{table*}

